\documentstyle[aps,eqsecnum,preprint,tighten,floats,epsf,rotate]{revtex}

\newbox\rotbox

\begin{document}
\draft

%%%%%%%%%%%%%%%%%%%%%%%%%%%%%%%%%%%%%%%%%%%%%%%%%%%%%%%%%%%%%%%%%%%%%%%%%%%%%%

\preprint{\vbox{\it 
                     \null\hfill\rm TRI-PP-96-35}\\\\}
%%%%%%%%%%%%%%%%%%%%%%%%%%%%%%%%%%%%%%%%%%%%%%%%%%%%%%%%%%%%%%%%%%%%%%%%%%%%%%

%
\title{CHANGE OF MIT BAG CONSTANT \\IN NUCLEAR MEDIUM
AND \\IMPLICATION FOR THE EMC EFFECT}
\author{Xuemin Jin\thanks{%
Address after September 1, 1996: Center for Theoretical Physics,
Laboratory for Nuclear Science and Department of Physics,
Massachusetts Institute of Technology, Cambridge, 
Massachusetts 02139, USA}
and B.K.~Jennings}
\address{TRIUMF, 4004 Wesbrook Mall\\ Vancouver, B.C,
                V6T 2A3, Canada} 
%
%\date{\today}
%
\maketitle

\begin{abstract}
  The modified quark-meson coupling model, which features a density dependent
  bag constant and bag radius in nuclear matter, is checked against the EMC
  effect within the framework of dynamical rescaling. Our emphasis is on the
  change in the average bag radius in nuclei, as evaluated in a local density
  approximation, and its implication for the rescaling parameter. We find that
  when the bag constant in nuclear matter is significantly reduced from its
  free-space value, the resulting rescaling parameter is in good agreement with
  that required to explain the observed depletion of the structure functions in
  the medium Bjorken $x$ region. Such a large reduction of the bag constant
  also implies large and canceling Lorentz scalar and vector potentials for the
  nucleon in nuclear matter which are comparable to those suggested by the
  relativistic nuclear phenomenology and finite-density QCD sum rules.
\end{abstract}
%
%\pacs{PACS numbers: 24.85+p; 25.30.Mr; 13.60.Hb; 12.39.Ba}

%%%%%%%%%%%%%%%%%%%%%%%%%%%%%%%%%%%%%%%%%%%%%%%%%%%%%%%%%%%%%%%%%%%%%%
\newpage
\narrowtext

\section{Introduction}
\label{intro}
While most nuclear models treat nucleons and mesons as the relevant
degrees of freedom for describing low- and medium-energy nuclear physics,
nuclear effects on nucleon structure functions, the EMC effect \cite{emc}, 
reveal the distortion of internal structure of the nucleon by the nuclear 
medium \cite{arneodo94}. To study this distortion, it is desirable 
to build models that incorporate the fundamental building blocks
of the nucleon, quark and gluon degrees of freedom, yet respect the
established theories based on hadronic degrees of freedom. Since the 
underlying theory of strong interactions, quantum chromodynamics (QCD),
is intractable at the nuclear physics energy scales, such models are
necessarily quite crude. 

The quark-meson coupling (QMC) model, proposed by Guichon \cite{guichon88}, 
provides a simple and attractive framework to incorporate the quark structure 
of the nucleon in the study of nuclear phenomena. In this model, nuclear
matter consists of non-overlapping MIT bags interacting through the
self-consistent exchange of mesons in the mean-field approximation, and the
mesons are coupled directly to the quarks inside the nucleon bags. This 
simple QMC model has been refined by including nucleon Fermi motion and 
the center-of-mass corrections to the bag energy \cite{fleck90} and 
applied to variety of nuclear physics problems
\cite{guichon88,fleck90,saito94,saito94a,saito92,guichon95,blunden96}.
(There have been several works that also discuss the quark effects in
nuclei, based on other effective models for the nucleon \cite{banerjee92}).

Recently, the present authors have pointed out that the assumption of 
fixing the MIT bag constant at its free-space value, adopted in the 
simple QMC model, is questionable and have modified the QMC model 
by allowing the bag constant to depend on the local density or sigma field.
\cite{jin96,jin96a,jin96b}. This modification can lead to the recovery
of relativistic nuclear phenomenology, in particular the large canceling 
isoscalar Lorentz scalar and vector potentials and hence the strong
spin-orbit force for the nucleon in nuclear matter. It is known that
such features are essential to the success of the relativistic nuclear
phenomenology \cite{wallace87}. However,  comparison to 
relativistic nuclear phenomenology \cite{wallace87} 
and finite-density QCD sum rules \cite{cohen95} suggests a large
reduction of the bag constant in nuclear matter.

In this paper, we examine the implications for the EMC effect of having the bag
constant decrease in the nuclear medium.  We shall be concerned only with the
so-called EMC effect region, i.e., the Bjorken $x$ region of $0.2 < x < 0.7$.
The most important feature in this region is that the structure function of a
bound nucleon is depleted with respect to that of a free nucleon and this
depletion is nucleus dependent \cite{arneodo94}. Since the structure functions
in the medium $x$ regime are dominated by the valence quark distributions, it
may be reasonable to investigate the EMC effect in a simple valence-quark
picture like the QMC model. (For smaller $x$, shadowing and antishadowing play
a dominant role, and for larger $x$, Fermi motion of the nucleon takes over
\cite{arneodo94}.)

Among many proposed theoretical explanations of the EMC effect
is the dynamical rescaling \cite{close83,jaffe84,close85,close85a,close88}. 
(Reviews of various proposed mechanisms can be found in 
Ref.~\cite{arneodo94}). This approach relies on having  the effective 
confinement size of quarks and gluons in a nucleus greater 
than that in a free nucleon \cite{jaffe83}. Such a change in 
confinement scale implies a reduction in the momentum carried by the
valence quarks and hence predicts, in the framework of perturbative QCD, 
that the structure function (in the EMC effect region) of a bound 
nucleon in nuclei can be related to that of a free nucleon by 
rescaling (in $Q^2$). The crucial input is the rescaling parameter,
$\xi_A(Q^2)$ [Eq.~(\ref{xi-def})], which is determined by the extent to 
which the confinement size changes from a free nucleon to a bound 
nucleon. This change was estimated in 
Refs.~\cite{close83,jaffe84,close85} by modeling the overlap of two nucleons, 
and it was found, for example, that the data can be explained if the 
confinement size in iron is $\sim 15\%$ larger than in an isolated nucleon. 
(The connections between the change of confinement scale and the EMC effect 
have also been discussed in Refs.~\cite{nachtmann84,deus84,hendry84,%
fredriksson84,cleymans85}).

Decreasing the MIT bag constant in the nuclear medium, as implemented in the
modified QMC model, implies a decrease of the bag pressure in nuclear
environment. This leads to an increase of the bag radius in nuclei relative to
its free-space value. Thus, the prediction of a change in the effective quark
confinement size emerges naturally in the modified QMC model. This change
yields a prediction for the rescaling parameter, which, in turn, gives rise to
predictions for the EMC effect in the framework of dynamical rescaling.

We use a local density approximation to evaluate the average bag radius 
in a nucleus. This radius is then used to determine the rescaling parameter.
We find that when the bag constant is significantly reduced in nuclear 
matter, e.g. $B/B_0 \sim 35-40\%$ at the nuclear matter saturation density, 
the predictions for the rescaling parameter are in good agreement with 
those required to explain the depletion of the structure function 
observed in a range of nuclei. Such a large reduction of the bag constant, 
as shown in previous works \cite{jin96,jin96a}, also implies large and 
canceling Lorentz scalar and vector potentials for the nucleon in nuclear 
matter which are comparable to those suggested by the relativistic nuclear 
phenomenology and finite-density QCD sum rules. This indicates that the 
reduction of bag constant and hence the increase of confinement size in 
nuclei may play important role in low- and medium-energy nuclear physics 
and the modified QMC model provides a useful framework to accommodate both 
the change of confinement size and the quark structure of the nucleon in 
describing nuclear phenomena.

The nuclear structure functions have been studied by Thomas and 
collaborators \cite{saito92} within the simple QMC model. (Similar 
study based on a soliton model for the in-medium nucleon was carried
out in Ref.~\cite{naar93}.) The technique assumed in these works is 
the same as the one used in calculating structure functions of a free 
nucleon from the MIT bag model \cite{jaffe83a}. We note that the resolution 
scale has been fixed at its value for free nucleon in these works. This 
is reasonable in the simple QMC model as the bag radius in nuclear matter 
only changes slightly compared to its free-space value. However, 
in the modified QMC model, the in-medium bag radius can be significantly
altered
and the approach used in Ref. \cite{saito92} cannot be adopted directly.
Our aim here is to simply study the implications of the change of 
confinement size in nuclei with respect to its free-space value for 
the EMC effect.

This paper is organized as follows: In Section~\ref{qmc-model} we
sketch the modified QMC model for nuclear matter and evaluate the 
average bag radius in finite nuclei by using a local density approximation. 
In Section~\ref{qscaling} we calculate the rescaling parameter from
the change of the average bag radius relative to the bag radius of
a free nucleon and discuss its implications for the EMC effect in 
the framework of dynamical rescaling. Further discussions are given 
in Section~\ref{discussion}. Section~\ref{conclusion} is a summary.
%%%%%%%%%%%%%%%%%%%%%%%%%%%%%%%%%%%%%%%%%%%%%%%%%%%%%%%%%%%%%%%%%%%%%%%%%%%%%%
%
\section{The modified QMC model and bag radius in nuclei}
\label{qmc-model}

In this section, we first give a brief introduction to the modified quark-meson
coupling model and then evaluate the average bag radius in nuclei by using a
local density approximation. In the modified quark-meson coupling model the
bag constant decreases and the nucleon bag swells when the nucleon is imbedded
in the nuclear medium. The reader is referred to Refs.
\cite{jin96,jin96a,jin96b} for further details and motivations for introducing
the medium modification of the bag constant.

%%%%%%%%%%%%%%%%%%%%%%%%%%%%%%%%%%%%%%%%%%%%%%%%%%%%%%%%%%%%%%%%%%%%%%%
\subsection{The modified quark-meson coupling model}

The QMC model depicts the nucleon in nuclear medium as a static spherical 
MIT bag in which quarks interact with the scalar and vector fields, 
$\overline{\sigma}$ and $\overline{\omega}$ \cite{guichon88}. These fields 
are treated as classical fields in the mean field approximation. The up and
down quark fields, $\psi_q(t, {\bf r})$, inside the nucleon bag then satisfies
the equation of motion: 
\begin{equation}
\left[i\,\rlap{/}\partial-(m_q^0-g_\sigma^q\, \overline{\sigma})
-g_\omega^q\, \overline{\omega}\,\gamma^0\right]\,\psi_q(t, {\bf r})=0\ ,
\label{eq-motion}
\end{equation}
where $m_q^0$ is the current quark mass, and $g_\sigma^q$ and
$g_\omega^q$ denote the quark-meson coupling constants.  For simplicity 
we will neglect isospin breaking and take $m_q^0=(m_u^0+m_d^0)/2 = 0$ 
hereafter. Inclusion of small current quark masses only yields numerically
small refinements.

The energy of a static bag consisting of three ground state quarks 
can be expressed as
\begin{equation}
E_{\rm bag}=3\, {\Omega_q\over R}-{Z\over R}
+{4\over 3}\,  \pi \, R^3\,  B\ ,
\label{ebag}
\end{equation}
where $\Omega_q\equiv \sqrt{y^2+(R\, m_q^*)^2}$, 
$m_q^*=m_q^0-g_\sigma^q\, \overline{\sigma}$, $R$ is the bag radius, 
$Z$ is a parameter which accounts for zero-point motion
and $B$ is the bag constant. The $y$ value is determined by the 
boundary condition at the bag surface, $j_0(y)=\beta_q\, j_1(y)$,
with $\beta_q = [(\Omega_q - R m^*_q)/(\Omega_q + R m^*_q)]^{1/2}$.
In the discussions to follow, we use $R_0$, $B_0$ and $Z_0$ to denote 
the corresponding bag parameters for the free nucleon. After the 
corrections of spurious center-of-mass motion in the bag, the effective 
mass of a nucleon bag at rest is taken to be\cite{fleck90,saito94}
\begin{equation}
M_N^*=\sqrt{E_{\rm bag}^2-\langle p_{\rm cm}^2\rangle}\ ,
\label{eff-mn}
\end{equation}
where $\langle p_{\rm cm}^2\rangle=\sum_q \langle p_q^2\rangle$ and
$\langle p_q^2\rangle$ is the expectation value of the quark momentum
squared, $(y/R)^2$. 

The equilibrium condition for the bag is obtained by 
minimizing the effective mass $M_N^*$ with respect to the bag radius
\begin{equation}
{\partial M_N^*\over \partial R} = 0\ .
\label{balance}
\end{equation}
In free space, one may fix $M_N$ at 
its experimental value $939$ MeV and use the equilibrium condition
to determine the bag parameters. For several choices of bag radius,
$R_0 = 0.6, 0.8, 1.0$ fm, the results for $B_0^{1/4}$ and $Z_0$
are $188.1, 157.5, 136.3$ MeV and $2.030,1.628,1.153$, respectively.
Finally, the scalar mean field is determined by the thermodynamic condition
\begin{equation}
\left({\partial\, E_{\rm tot}\over 
\partial\, \overline{\sigma} }\right)_{R,\rho_N} = 0\ .
\label{thermal}
\end{equation}

The simple QMC model assumes that both $Z$ and $B$ are independent of 
density \cite{guichon88,fleck90,saito94}. Such an assumption is 
questionable \cite{jin96,jin96a,jin96b}. We have proposed two models 
for the modification of the bag constant.\footnote{%
In principle, the parameter $Z$ may also be modified
in the nuclear medium. However, it is unclear how $Z$ changes with the density.
Here we assume that the medium modification of $Z$ is small at low and moderate 
densities and take $Z=Z_0$. Recently, Blunden and Miller \cite{blunden96} have
considered a density dependent $Z$. However, it is found that for
reasonable parameter ranges changing $Z$ has little effect and tends to
make the model worse.} The direct coupling model \cite{jin96a} (model-I) 
invokes a direct coupling of the bag constant to the scalar meson field 
\begin{equation}
{B\over B_0} = \left[ 1
- g_\sigma^B\, {4 \over \delta} {\overline{\sigma}\over M_N} \right]^\delta\ ,
\label{an-dir}
\end{equation}
where $g_\sigma^B$ and $\delta$ are real positive parameters. 
In the limit of $\delta \rightarrow \infty$, Eq.~(\ref{an-dir})
reduces to an exponential form, $B/B_0 = e^{-4 g_\sigma^B 
\overline{\sigma}/M_N}$, with a single parameter $g_\sigma^B$.
The scaling model \cite{jin96,jin96a} (model-II) relates the in-medium 
bag constant directly to the in-medium nucleon mass $M_N^*$
\begin{equation}
{B\over B_0} = \left[ M_N^*\over M_N \right]^\kappa\ ,
\label{an-br}
\end{equation}
where $\kappa$ is a real positive parameter. 
%%%%%%%%%%%%%%%%%%%%%%%%%%%%%%%%
\begin{figure}[t]
\begin{center}
\epsfysize=11.7truecm
\leavevmode
\setbox\rotbox=\vbox{\epsfbox{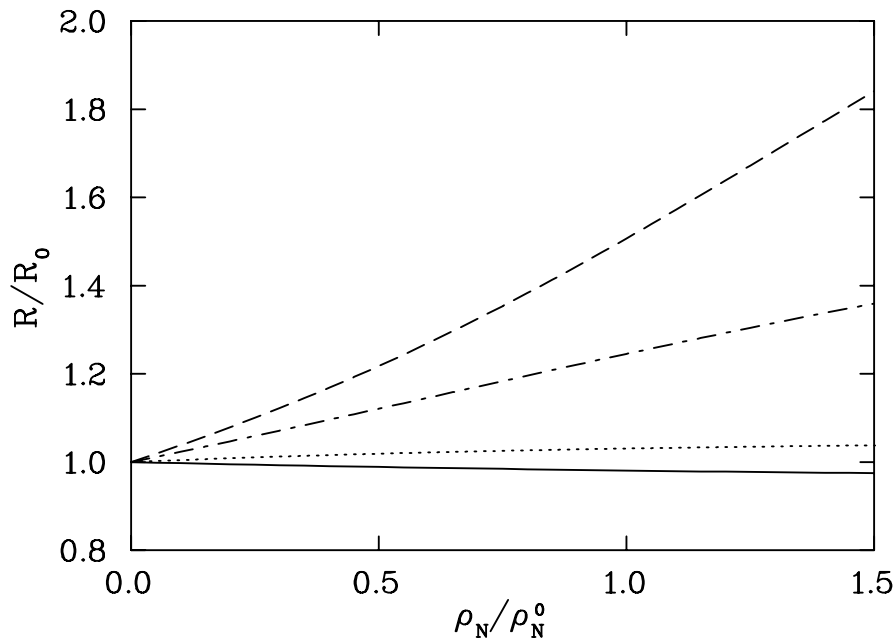}}\rotl\rotbox
\end{center}
\caption{Result for the ratio $R/R_0$ as a function of the medium density, 
with $\delta = 4$ and $R_0=0.6$ fm. Here model-I Eq.~(\protect\ref{an-dir}) 
is used for the in-medium bag constant. The couplings $g_\sigma^B$ and 
$g^q_\sigma$ are adjusted to fit the nuclear matter binding energy and 
the other parameters are the same as used in Ref. \protect\cite{jin96a}. 
The four curves correspond to $g_\sigma^q = 1.0$ (long-dashed), 2.0 (dot-dashed), 
4.0 (dotted), and 5.309 (solid), respectively.}
\label{fig-1}
\end{figure}
%%%%%%%%%%%%%%%%%%%%%

One notices that both Eqs.~(\ref{an-dir}) and ({\ref{an-br}) give rise
to a reduction of the bag constant in nuclear medium relative to its
free-space value. While the scaling model is characterized by a single
free parameter $\kappa$, it leads to a complicated and implicit 
relation between the bag constant and the scalar mean field. On the other
hand, the direct coupling model features a straightforward coupling
between the bag constant and the scalar mean field, which, however, 
introduces two free parameters, $g_\sigma^B$ and $\delta$. 
Various couplings and other parameters can be adjusted to reproduce the 
nuclear matter binding energy ($-16$ MeV) at the saturation density 
($\rho_{\rm N}^0=$0.17 fm$^{-3}$). The resulting coupling constants 
and nuclear matter results have been given in Refs.~\cite{jin96,jin96a}. 

Here we are interested in the modification of the bag radius in nuclear medium. 
In the usual QMC model, the bag radius decreases slightly ($\sim 1\%$) and 
the quark root-mean-square (RMS) radius increases slightly ($\sim 1\%$) in 
saturated nuclear matter with respect to their free-space values 
\cite{guichon88,saito94}. When the bag constant drops relative 
to its free-space value, the bag pressure decreases and hence the bag radius 
increases in the medium. It has been found in Refs.~\cite{jin96a,jin96b} 
that if the reduction of the bag constant is significant 
(e.g. $B/B_0 \sim 35-40\%$), the bag radius in saturated nuclear matter 
is $25 - 30\%$ larger than its free-space value. This result is essentially 
determined by the value of $B/B_0$ at $\rho^0_N$ and largely insensitive to 
the model used for the in-medium bag constant. (The quark RMS radius also 
increases with density, with essentially the same rate as for the bag radius.) 
As an example, we have plotted in Fig.~\ref{fig-1} the resulting 
bag radius as a function of $\rho_N$, with model-I, Eq.~(\ref{an-dir}), for 
the in-medium bag constant. The corresponding bag constant is shown in 
Fig.~\ref{fig-2}. The results from model-II, Eq.~(\ref{an-br}), can be found 
in Ref.~\cite{jin96}.

%%%%%%%%%%%%%%%%%%%%%%%%%%%%%%%%
\begin{figure}[t]
\begin{center}
\epsfysize=11.7truecm
\leavevmode
\setbox\rotbox=\vbox{\epsfbox{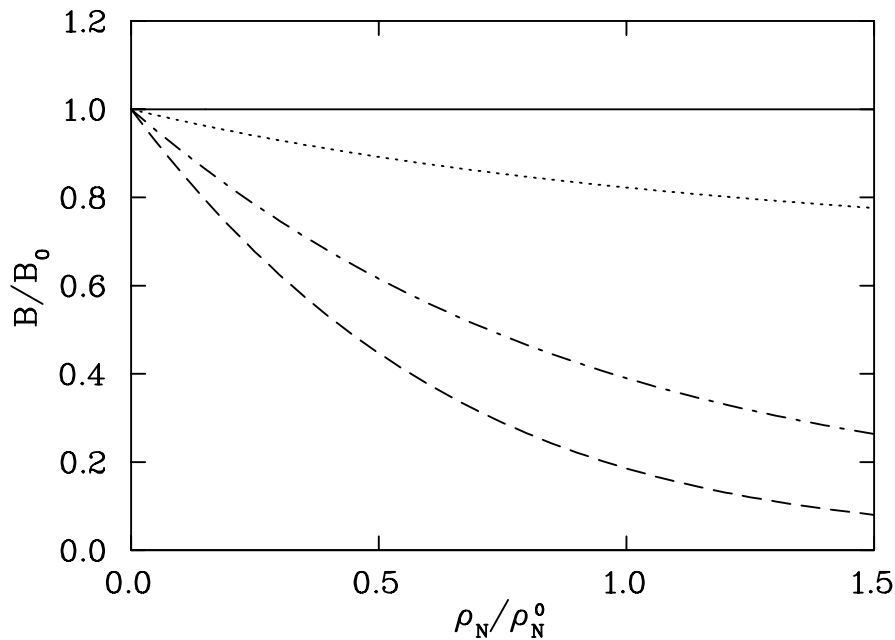}}\rotl\rotbox
\end{center}
\caption{Corresponding result for the ratio $B/B_0$ as a function of the medium 
density. The same parameters are used as in Fig.~\protect\ref{fig-1}. The four 
curves correspond to $g_\sigma^q = 1.0$ (long-dashed), 2.0 (dot-dashed), 
4.0 (dotted), and 5.309 (solid), respectively.}
\label{fig-2}
\end{figure}
%
%%%%%%%%%%%%%%%%%%%%%%%%%%%%%%%%%%%%%%%%%%%%%%%%%%%%%%%%%%%%%%%%%%%%%%%%%%%%%
\subsection{Bag radius in finite nuclei}

With the density dependence of the bag radius obtained from the modified 
QMC model, we are ready to evaluate the average bag radius in a finite 
nucleus, $A$, in the local density approximation:
\begin{equation}
\overline{R}_A \equiv {\int d^3 r \, R[\rho_A(r)]\, \rho_A(r)\over
 \int d^3 r \, \rho_A(r)} ,
\label{lda}
\end{equation}
where $\rho_A(r)$ is the density distribution of the nucleus $A$
and $R[\rho_A(r)]$ denotes the bag radius at the density
$\rho_A(r)$. Here we adopt the phenomenological fits in a form of the 
Woods-Saxon type function for the density distribution 
$\rho_A(r)$ \cite{barrett77}
\begin{equation}
\rho_A(r) = {\overline{\rho}_A \over 
1+\mbox{exp}[(r - {\cal R}_A)/a]} \ .
\label{rhoA}
\end{equation}
Here the three parameters, $\overline{\rho}_A$, ${\cal R}_A$ and $a$, are
used to fit shapes of nuclei. Their values for various nuclei \cite{barrett77}
are listed in Table \ref{tab-1}. 
%%%%%%%%%%%%
\widetext
\begin{table}[t]
\caption{Parameters $\overline{\rho}_A$, ${\cal R}_A$, and $a$ in
Eq. (\protect{\ref{rhoA}}) to fit the shapes of nuclei with $A=$ 12, 
20, 27, 56, 63, 107, 118, 197, and 208 (from Ref.~\protect\cite{barrett77}).}
\label{tab-1}
\begin{tabular}{lccc}
Nucleus & $\overline{\rho}_A$ (fm$^{-3}$) & ${\cal R}_A$ (fm) & $a$ (fm) \\
\tableline
$^{12}$C     &  0.1708    &   2.240   &     0.500\\
$^{20}$Ne    &  0.1628    &   2.740   &     0.569\\
$^{27}$Al    &  0.1738    &   3.070   &     0.519\\
$^{56}$Fe    &  0.1760    &   3.980   &     0.569\\
$^{63}$Cu    &  0.1674    &   4.218   &     0.596\\
$^{107}$Ag   &  0.1566    &   5.299   &     0.523\\
$^{118}$Sn   &  0.1607    &   5.412   &     0.560\\
$^{197}$Au   &  0.1694    &   6.380   &     0.535\\
$^{208}$Pb   &  0.1600    &   6.624   &     0.549\\
\end{tabular}
\end{table}
%%%%%%%%%%%%%%%%%%%%%%%%%%%%%%%%

Figure~\ref{fig-3} shows the resulting ratio $\overline{R}_A/R_0$ 
from model-I Eq.~(\ref{an-dir}) as a function of $g_\sigma^q$ for 
different nuclei, with $\delta = 4$ and $R_0 = 0.6$ fm. Here the couplings 
$g_\sigma^B$ and $g^q_\omega$ are adjusted to fit the nuclear matter binding 
energy. The case $g_\sigma^q = 0$ ($\delta =4$) corresponds to QHD-I 
(but with density dependent bag radius) \cite{jin96a}. 
The reduction of $B$ in this case is large ($B/B_0 \sim 10\%$). 
On the other hand, $g_\sigma^q \simeq 5.309$ gives the usual QMC model,
where the bag constant is independent of density (i.e., $B/B_0 = 1$). When
$g_\sigma^q > 5.309$, the in-medium bag constant increases instead
of decreases relative to $B_0$ (i.e., $B/B_0 > 1)$. 
%%%%%%%%%%%%%%%%%%%%%%%%%%%%%%%%
\begin{figure}[t]
\begin{center}
\epsfysize=11.7truecm
\leavevmode
\setbox\rotbox=\vbox{\epsfbox{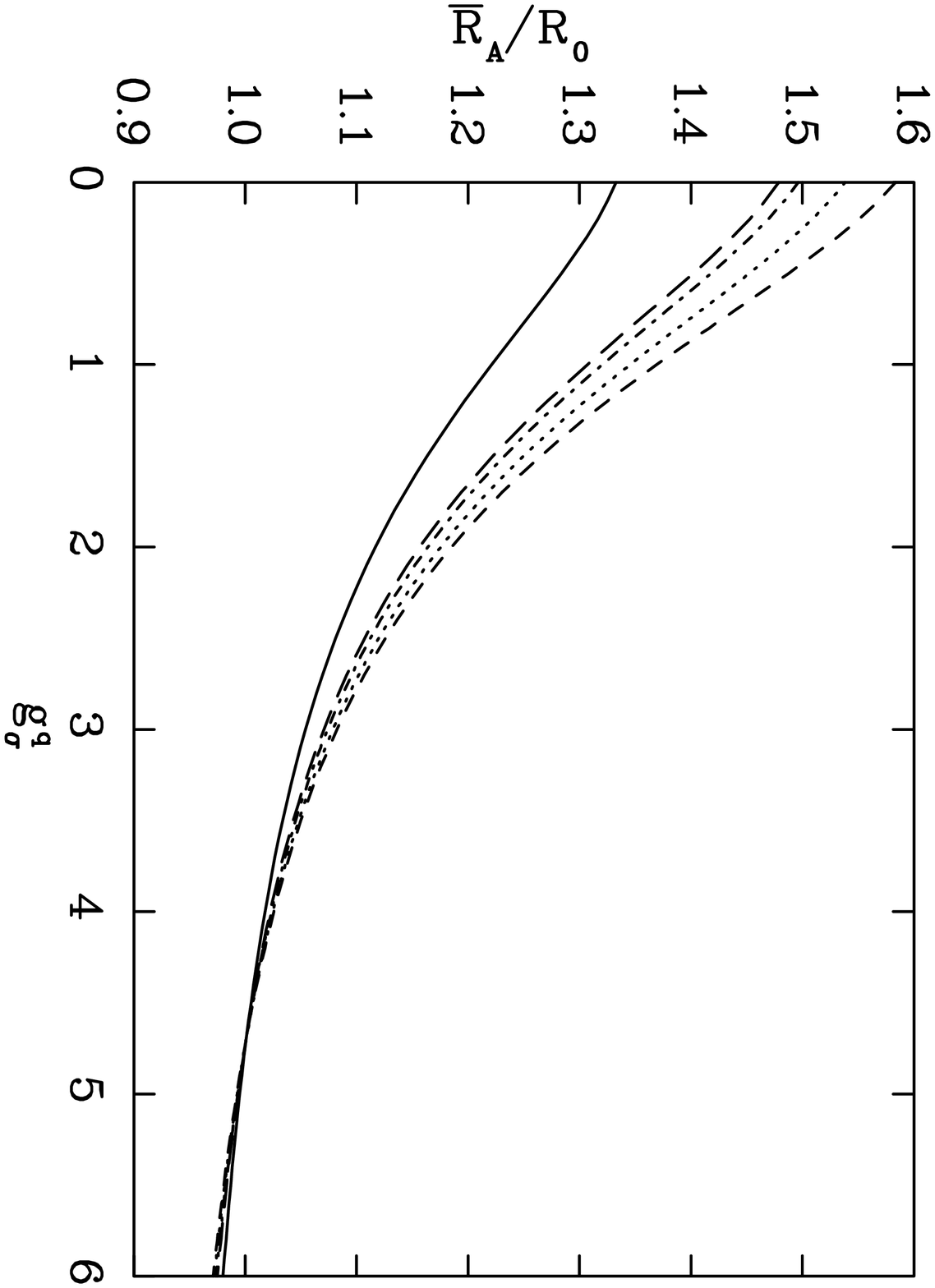}}\rotl\rotbox
\end{center}
\caption{Result for the ratio $\overline{R}_A/R_0$ as a function of 
$g_\sigma$, with $\delta = 4$ and $R_0=0.8$ fm. Here model-I 
Eq.~(\protect\ref{an-dir}) is adopted. The five curves correspond 
to $A=12$ (solid), 56 (long-dashed), 118 (dot-dashed), 197 (short-dashed), 
and 208 (dotted), respectively.}
\label{fig-3}
\end{figure}
%%%%%%%%%%%%%%%%%%

It can be seen from Fig.~\ref{fig-3} that for small $g_\sigma^q$, 
$\overline{R}_A$ changes significantly with respect to $R_0$. 
As $g^q_\sigma$ increases, the ratio $\overline{R}_A/R_0$ decreases 
quickly resulting from the increase of $B/B_0$. The $A$ dependence of 
$\overline{R}_A/R_0$ is strong for light nuclei. As $A$ becomes large,
the $A$ dependence weakens and the ratio $\overline{R}_A/R_0$ appears
to saturate. We also find that for a given $g_\sigma^q$, 
increasing $\delta$ leads to the increase of $B/B_0$ and hence the 
decrease of $\overline{R}_A/R_0$, and for fixed $g_\sigma$ and $\delta$, 
the results are insensitive to the choice of $R_0$.

The result from model-II Eq.~(\ref{an-br}) is illustrated in Fig.~\ref{fig-4}, 
where the ratio $\overline{R}_A/R_0$ as a function of $\kappa$ is plotted 
with $R_0 = 0.6$ fm. Here the quark-meson couplings $g_\sigma^q$ and 
$g_\omega^q$ are chosen to reproduce the nuclear matter binding energy. 
The case $\kappa = 0$ corresponds to the usual QMC model. For small $\kappa$ 
values ($\kappa < 1.2$), the ratio $\overline{R}_A/R_0$ is close to unity.
As $\kappa$ gets larger, $\overline{R}_A/R_0$ increases rapidly. This 
is due to the decrease of $B/B_0$ with increasing $\kappa$. The ratio
$\overline{R}_A/R_0$ in this case has similar $A$ dependence and sensitivity 
to $R_0$ as in model-I.
%%%%%%%%%%%%%%%%%%%%%%%%%
\begin{figure}[t]
\begin{center}
\epsfysize=11.7truecm
\leavevmode
\setbox\rotbox=\vbox{\epsfbox{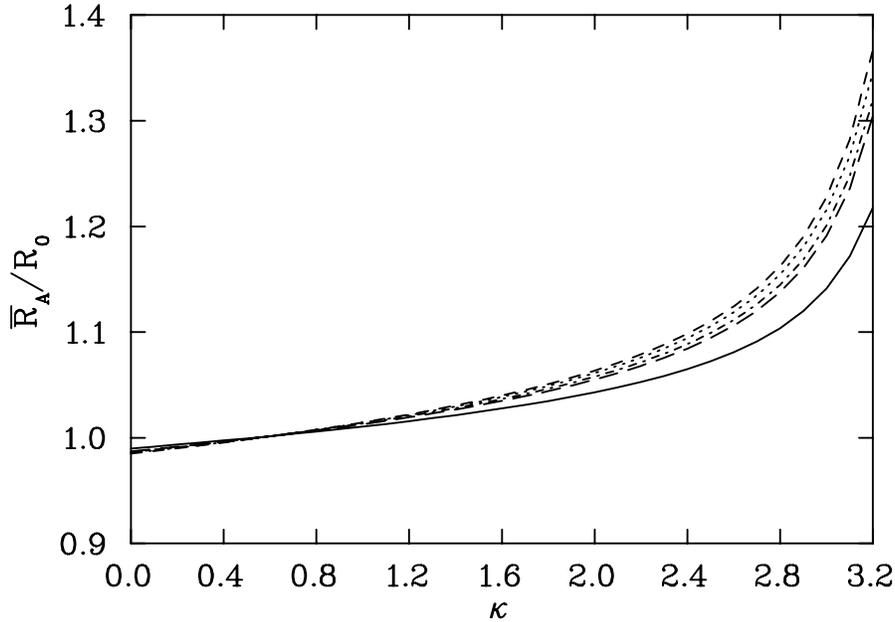}}\rotl\rotbox
\end{center}
\caption{Result for the ratio $\overline{R}_A/R_0$ as a function of 
$\kappa$, with $R_0=0.6$ fm. Here model-II Eq.~(\protect\ref{an-br})
is adopted. The five curves correspond to $A=12$ (solid), 56 (long-dashed), 
118 (dot-dashed), 197 (short-dashed), and 208 (dotted), respectively.}
\label{fig-4}
\end{figure}
%%%%%%%%%%%%%%%%%%%%%%%%%%

Wo note that the results for $\overline{R}_A/R_0$ are largely 
controlled by the change of the bag constant in nuclear medium with 
respect to its free-space value. To illustrate this point,
we have listed the values of $\overline{R}_A/R_0$ for different 
nuclei in Table \ref{tab-2}, with the ratio $B/B_0$ fixed at
$B/B_0 =  40\%$ (at $\rho_N = \rho_N^0$). One can see that the results 
are fairly model independent and the sensitivity to $R_0$ is very samll 
($\sim 1\%$). When $R_0$ increases, the ratio $\overline{R}_A/R_0$
increases slightly. This variation with $R_0$ can be compensated
by tuning down the ratio $B/B_0$ slightly. We have also tested the
sensitivity of model-I results to the $\delta$ value and found that the 
results are insensitive to $\delta$. (In the limit 
of $\delta\rightarrow \infty$, $B/B_0$ at $\rho_N = \rho_N^0$ is 
always larger than $40\%$ for positive $g_\sigma^q$ value).
%%%%%%%%%%%%%%%%%%%%%%%%%%%
%
\begin{table}[t]
\caption{Predictions for the ratio $\overline{R}_A/R_0$, with $B/B_0$
fixed to $B/B_0 = 40\%$ (at $\rho_N =\rho_N^0$).}
\label{tab-2}
\begin{tabular}{lcccccc}
Nucleus    &\multicolumn{3}{c}{$\overline{R}_A/R_0$ (Model-I with $\delta = 4$)}
&\multicolumn{3}{c}{$\overline{R}_A/R_0$ (Model-II)} \\
 & $R_0=0.6$ fm & 0.8 fm& 1.0 fm& $R_0=0.6$ fm & 0.8 fm& 1.0 fm \\
\tableline
$^{12}$C     &1.113& 1.119& 1.121& 1.115& 1.121& 1.123\\
$^{20}$Ne    &1.113& 1.118& 1.121& 1.115& 1.121& 1.123\\
$^{27}$Al    &1.137& 1.143& 1.146& 1.139& 1.146& 1.148\\
$^{56}$Fe    &1.152& 1.159& 1.162& 1.154& 1.161& 1.164\\
$^{63}$Cu    &1.145& 1.152& 1.155& 1.147& 1.154& 1.157\\
$^{107}$Ag   &1.157& 1.165& 1.168& 1.160& 1.168& 1.170\\
$^{118}$Sn   &1.159& 1.166& 1.169& 1.162& 1.169& 1.172\\
$^{197}$Au   &1.179& 1.188& 1.191& 1.182& 1.190& 1.194\\
$^{208}$Pb   &1.170& 1.178& 1.181& 1.173& 1.181& 1.184\\
\end{tabular}
\end{table}
%
%%%%%%%%%%%%%%%%%%%%%%%%%%%%%%%%%%%%%%%%%%%%%%%%%%%%%%%%%%%%%%%%%%%
\section{Implication for the EMC effect in dynamical rescaling}
\label{qscaling}

In this section, we calculate the rescaling parameter from the change 
of the average bag radius in nuclei obtained in previous section and 
discuss its implications for the EMC effect in the dynamical rescaling
approach. We refer the reader to 
Refs. \cite{close83,jaffe84,close85,close85a,close88}
for further details and physical motivations of the approach. 

The dynamical rescaling is based on that the confinement size for quarks
in nuclei is larger than in free nucleons. This then leads to a rescaling 
relation \cite{close83,jaffe84,close85,close85a,close88}
\begin{equation}
F^A_2(x,Q^2) = F^N_2(x, \xi_A Q^2) \ ,
\label{res-re}
\end{equation}
which connects the structure function (per nucleon) in the nucleus 
$A$, $F^A_2(x,Q^2)$, to the nucleon structure function in free space,
$F^N_2(x,Q^2)$, where $x$ is the Bjorken variable and $Q^2$ is the probing 
momentum. Here the parameter $\xi_A(Q^2)$ is called rescaling parameter, which 
can be related to the ratio of the quark confinement scale in the nucleus 
$A$ and that in the nucleon \cite{close85a}
\begin{equation}
\xi_A (Q^2) = \left[\left({\overline{R}_A\over 
R_0}\right)^2\right]^{\alpha_s(\mu^2)/\alpha_s(Q^2)}\ ,
\label{xi-def}
\end{equation}
with
\begin{equation}
\alpha_s(\mu^2)/\alpha_s(Q^2) = 
\ln(Q^2/\Lambda^2_{\rm QCD})/\ln(\mu^2/\Lambda^2_{\rm QCD})\ ,
\label{alph-r}
\end{equation}
where $\Lambda_{\rm QCD}$ is the QCD scale parameter. Here we follow
Ref.~\cite{close85a} and take $\Lambda_{\rm QCD} = 0.25$ GeV and
$\mu^2 = 0.66$ GeV$^2$. In Eq.~(\ref{xi-def}) we have 
identified $\overline{R}_A$ and $R_0$ as the quark confinement sizes 
in the nucleus $A$ and in the free nucleon, respectively. The relation
Eq.~(\ref{res-re}) can also be generalized to any pair of nuclei, with
$\xi_A$ replaced by $\xi_{AA^\prime}$ and $R_0$ replaced by 
$\overline{R}_{A^\prime}$. 

The rescaling relation Eq.~(\ref{res-re}) can be understood by considering 
the moments of the nuclear structure function defined by
$M^A_n(Q^2) \equiv \int^A_0 dx\, x^{n-2} F^A_2(x,Q^2)$. If one knows the 
moment $M^A_n(Q^2)$ at some initial value of $Q^2$, say $\mu^2_A$ which 
can be regarded as the low-momentum cut-off for radiative gluons, then 
it is calculable for all $Q^2 > \mu^2_A$. The idea of dynamical rescaling 
is that, since in perturbative QCD the target dependence resides in the 
nonperturbative matrix elements, the scale $\mu^2_A$ may be target dependent 
and independent of $n$, such that $M^A_n(\mu^2_A) = M^N_n(\mu^2_N)$. Then, 
one finds for $Q^2 > \mu^2_A$, $M^A_n(Q^2) = M_n^N (\xi_A Q^2)$, which 
implies the relation given by Eq.~(\ref{res-re}). Further details and 
discussions can be found in Ref.~\cite{close83,jaffe84,close85,close85a,close88}.

The parameter $\xi_A(Q^2)$ is the most important input in the rescaling.
Feeding the ratio $\overline{R}_A/R_0$ obtained in previous 
section into Eq.~(\ref{xi-def}), one obtains the predictions of the
modified QMC model for $\xi_A(Q^2)$. The resulting values at $Q^2 = 20$ 
GeV$^2$ are listed in Table~\ref{tab-3}, where the values of $\overline{R}_A/R_0$
given in Table~\ref{tab-2} (with $R_0 =0.6$ fm) are used. The results 
of Ref.~\cite{close85} are also given for comparison. One can see that 
the predictions for $\xi_A(Q^2)$ agree well with the results of 
Ref.~\cite{close85}. (For larger $R_0$, very similar results can be 
obtained with slightly smaller values of $B/B_0$.) In particular, for 
iron the predicted $\xi_A\simeq 2$ and $\overline{R}_A/R_0\simeq 1.15$ are 
essentially identical to those required to explain the experimental 
data \cite{close85}. The $A$ dependence of $\xi_A(Q^2)$ is somewhat 
weaker than that found in Ref.~\cite{close85}. This, however, has only 
small impact on the predictions for the EMC effect as the $Q^2$ dependence 
of the structure function is only logarithmic in the context of perturbative QCD.
%%%%%%%%%%%%%%%%%%%%%%%%%%%
\begin{table}[t]
\caption{Predictions for the rescaling parameter $\xi_A(Q^2)$ at $Q^2 = 20$ GeV$^2$.
Here the values of $\overline{R}_A/R_0$ with $R_0 = 0.6$ fm listed in 
Table~\protect\ref{tab-2} have been used. The last column gives the results
of Ref.~\protect\cite{close85} (using the Reid soft-core version of the 
correlation function).}
\label{tab-3}
\begin{tabular}{lccc}
Nucleus    &$\xi(Q^2)$ (Model-I with $\delta = 4$)& $\xi(Q^2)$ (Model-II)&
$\xi(Q^2)$ (Ref. \protect\cite{close85})\\
\tableline
$^{12}$C     &1.69& 1.70& 1.60\\
$^{20}$Ne    &1.69& 1.70& 1.60\\
$^{27}$Al    &1.88& 1.89& 1.89\\
$^{56}$Fe    &2.00& 2.02& 2.02\\
$^{63}$Cu    &1.95& 1.96& 2.02\\
$^{107}$Ag   &2.04& 2.07& 2.17\\
$^{118}$Sn   &2.06& 2.09& 2.24\\
$^{197}$Au   &2.24& 2.27& 2.46\\
$^{208}$Pb   &2.16& 2.18& 2.37\\
\end{tabular}
\end{table}
%%%%%%%%%%%%%%%%%%%%%%%%%%%%%%%%%%%%%%

It has been demonstrated in Refs.~\cite{close83,jaffe84,close85,close85a,close88}
that the values of rescaling parameter listed in the last column of Table~\ref{tab-3}
give predictions for the EMC effect in the medium $x$ region which are in excellent 
agreement with the experiment data. Since our predictions for $\xi_A(Q^2)$
are very close to those values, we expect that the resulting predictions for 
the EMC effect will also be very similar. Here we shall not repeat
the analysis. The interested readers can find the comparison between
the predictions of the dynamical rescaling and the experimental data
in Refs.~\cite{close83,jaffe84,close85,close85a,close88}.
Thus, when the bag constant drops significantly in nuclear matter,
$B/B_0 \simeq 35-40\%$, the predictions of the modified QMC model
for the change of average bag radius in nuclei relative the bag radius 
of an isolated nucleon can lead to satisfying explanation of the 
EMC effect in the medium $x$ region. 
%%%%%%%%%%%%%%%%%%%%%%%%%%%%%%%%%%%%%%%%%%%%%%%%%%%%%%%%%%%%%%%%%%%%
\section{Discussion}
\label{discussion}

As pointed out in Refs.~\cite{jin96,jin96a,jin96b}, a significant reduction 
of the bag constant also implies large potentials for the nucleon 
in nuclear matter. In particular, when $B/B_0 \simeq 35-40\%$ at 
$\rho_N = \rho_N^0$, we find \cite{d-choice}
\begin{equation}
M^*_N / M_N \sim 0.72 \ ,{\hspace*{1cm}
U_{\rm v} / M_N \sim 0.21 \ ,}
\label{try-result}
\end{equation}
where $U_{\rm v}\equiv 3 \, g_\omega^q \, \overline{\omega}$. Since the 
equivalent scalar and vector potentials appearing in the wave equation 
for a point-like nucleon are essentially $M^*_N-M_N$ and $U_{\rm v}$ 
\cite{guichon95,blunden96}, Eq.~(\ref{try-result}) shows large and canceling 
scalar and vector potentials for the nucleon in nuclear matter, which are 
comparable to those suggested by the relativistic nuclear phenomenology 
\cite{wallace87} and finite-density QCD sum rules \cite{cohen95}. 

Such a large reduction of the bag constant is not entirely 
unexpected. If one adopts the scaling ansatz advocated by Brown and Rho
\cite{brown91}, the in-medium bag constant scales like
\cite{adami93}, $B/B_0 \simeq \Phi^4$, where $\Phi$ denotes 
the universal scaling, $\Phi\simeq m_\rho^*/m_\rho \simeq f^*_\pi/f_\pi \cdots$,
which is density dependent. Here, the ``starred'' quantities refer to
the corresponding in-medium quantities. Taking the result
for $m_\rho^*/m_\rho$ from the most recent finite-density QCD sum-rule
analysis \cite{jin95}, one finds $\Phi\simeq m_\rho^*/m_\rho\sim 0.78$
at the saturation density, which gives rise to $B/B_0\simeq \Phi^4\sim
0.36$. There are, however, some caveats concerning this estimate which have 
been discussed in detail in Ref.~\cite{jin96a}.

Moreover, the dropping bag constant also leads to a swelling nucleon in
nuclear environment. This has important implications for 
various nuclear physics issues which have been discussed extensively in the 
literature~\cite{noble81,celenza84,sick85,brown88,brown89,soyeur93}.
In the modified QMC model, the nucleon swelling is also reflected 
in the outstretched quark wave functions (see Ref.~\cite{jin96a}).
This is supported by the studies of finite-density QCD sum 
rules~\cite{jin94} and other studies of the modification
of internal structure of the composite nucleon~\cite{frank95}.
It was found in these studies that there is a sizable reduction for
the nucleonic wave function at the origin in nuclear medium relative 
to that in free space, which is an indication of outstretched wave
function and nucleon swelling.

Therefore, the reduction of the bag constant in nuclear medium 
may play important role in low- and medium-energy nuclear physics. In 
particular, this reduction effectively introduces a new source of attraction 
for the nucleon which needs to be compensated with additional vector strength. 
On the other hand, the dropping bag constant also predicts the increase 
of quark confinement scale in nuclei which leads to the reduction of the 
momentum carried by the valence quarks and hence the depletion of the 
structure function in the medium $x$ region. It is satisfying to find 
the mutual consistency and connection between the large nucleon potentials 
and the EMC effect. Thus, the modified QMC model provides a simple 
and useful framework for describing nuclear phenomena, which incorporates 
the quark structure of the nucleon, yet respect the established relativistic 
nuclear phenomenology based on point-like nucleons and mesons. 

While it is attributed to the overlapping effect between two nucleons 
in Refs.~\cite{close83,jaffe84,close85,close85a,close88}, the change of 
quark confinement scale in nuclei results from the dropping bag constant 
in nuclear medium in the modified QMC model. The fact that the two approach 
give very similar predictions may imply that they describe similar physics. 
Our view is that the decrease of the bag constant in nuclear medium 
(through the coupling to the scalar mean field) and the resulting change 
of confinement size in nuclei effectively parametrize the physics of the 
nucleon overlapping effect and/or other more complicated nuclear dynamics. 
In this sense, the modified QMC model can be reconciled with the traditional 
picture of a nucleus as a collection of bound nucleons and the traditional
description of nuclear physics in terms of hadronic degrees of freedom.

We note that other effects such as nuclear binding and Fermi motion also 
contribute to the EMC effect in the medium $x$ regime.  These effects should
be applied in addition to the predictions of the dynamical rescaling if one 
is to fit the observed data. The inclusion of these effects may alter the 
phenomenological success obtained from the dynamical rescaling alone. 
For example, it is found in Ref.~\cite{li88} that the nuclear binding and 
Fermi motion in the conventional model of the nucleons account for about 
$20\%$ of the EMC effect in the medium $x$ region (see, however, 
Ref.~\cite{bickerstaff87}). Consequently, the change of quark confinement 
size in nuclei required to explain the data is somewhat smaller than that 
discussed in the present paper. 

However, the authors of Ref.~\cite{bickerstaff86} have claimed that dynamical 
rescaling mimics binding effects in the conventional model and it was argued
in Ref.~\cite{close88} that the nuclear convolution models \cite{bickerstaff89} 
and dynamical rescaling provide alternative not different explanations of 
the EMC effect. This is consistent with our view of regarding the reduction 
of the bag constant and change of confinement size in nuclei as simple 
parametrization of detailed nuclear dynamics. Clearly, further study is
needed to clarify whether the dynamical rescaling and the conventional 
approach are really independent or they describe the same physics.

Finally, it has been emphasized previously \cite{jin96a} that the QMC model 
is only a simple extension of QHD, where the exchanging mesons are treated 
as classical fields in the mean-field approximation. The explicit quark 
structures of the mesons should also be included and the physics beyond 
the mean-field approximation should be considered in a more consistent 
treatment. It is also known that both the MIT bag model and the QHD are 
not compatible with the chiral symmetry. To improve this situation, one 
may use a chiral version of the bag model and a relativistic hadronic model
consistent with the chiral symmetry.

%%%%%%%%%%%%%%%%%%%%%%%%%%%%%%%%%%%%%%%%%%%%%%%%%%%%%%%%%%%%%%%%%%%
\section{Summary}
\label{conclusion}

In this paper, we have confronted the modified quark-meson coupling
model developed previously with the EMC effect within the dynamical
rescaling approach. We used a local density approximation to evaluate
the average bag radius in nuclei and determined the rescaling parameter 
from the change of the average bag radius in nuclei relative to the 
radius of a free nucleon bag. 

When the bag constant is reduced significantly in nuclear matter, 
$B/B_0 \simeq 35-40\%$ at the nuclear matter saturation density, 
the predicted values for the rescaling parameter are in good agreement 
with those required to explain the EMC effect in the medium $x$ region. 
This result is largely independent of how the in-medium bag constant is 
modeled and is very stable against variations of various model parameters.
Such a significant reduction of the bag constant also leads to large and 
canceling scalar and vector potentials for the nucleon in nuclear 
matter which are consistent with those suggested by the relativistic
nuclear phenomenology and finite-density QCD sum rules. 

Since the predictions of the modified QMC model for the change of the 
average bag radius (quark confinement size) in nuclei are very similar 
to those obtained from modeling the overlap between two nucleons, it seems 
plausible that the reduction of the bag constant (through the coupling of 
the bag constant to the scalar mean field) effectively reflects the physics 
of nucleon overlapping and/or other detailed and complicated nuclear dynamics.

To conclude, the modified QMC model provides a useful framework for 
describing nuclear phenomena, in which the decrease of the bag 
constant in nuclear environment plays an important role. We have seen
that the model gives consistent predictions for large nucleon potentials 
in nuclear matter and the EMC effect. We look forward to further vindication 
of this model in addressing other nuclear physics problems.
%%%%%%%%%%%%%%%%%%%%%%%%%%%%%%%%%%%%%%%%%%%%%%%%%%%%%%%%%%%%%%%%%%%
\acknowledgments
This work was supported by the Natural Sciences and Engineering 
Research Council of Canada.
%%%%%%%%%%%%%%%%%%%%%%%%%%%%%%%%%%%%%%%%%%%%%%%%%%%%%%%%%%%%%%%%%%%%%

%
%%%%%%%%%%%%%%%%%%%%%%%%%%%%%%%%%%%%%%%%%%%%%%%%%%%%%%%%%%%%%%%%%%%%


\begin{references}
\bibitem{d-choice}In the direct coupling model, the parameter $\delta$
controls the non-linearities in $B(\overline{\sigma})$ and
$M_N^*(\overline{\sigma})$. The results listed in 
Eq.~(\protect\ref{try-result}) correspond to relatively small
$\delta$ values ($\delta \simeq 4-8$). As $\delta$ becomes large,
$M^*_N/M_N$ increases and $U_{\rm v}/M_N$ decreases. However, large
$\delta$ values gives very complicated relation between $B$ and the
scalar mean field.
\end{references}
\end{document}